\documentclass{article}
\usepackage{slashed,feynmf,epsfig,amsmath,amssymb,enumitem}
\usepackage[papersize={8.5in,11in}]{geometry}
\renewcommand{\vec}[1]{\boldsymbol{\mathrm{#1}}}
\usepackage{graphicx}
\begin{document}
\title{Cosmological Evidence for Modified Gravity (MOG)}
\author{J. W. Moffat\\~\\
Perimeter Institute for Theoretical Physics, Waterloo, Ontario N2L 2Y5, Canada\\
and\\
Department of Physics and Astronomy, University of Waterloo, Waterloo,\\
Ontario N2L 3G1, Canada}
\maketitle




\begin{abstract}
Deviations from the standard $\Lambda$CDM model motivate an interpretation of early universe cosmology using the Scalar-Tensor-Vector-Gravity (STVG) theory. A constraint analysis carried out by Valentino, Melchiorri and Silk, revealed deviations from the growth of structure predicted by General Relativity, and a lensing anomaly in the angular CMB power spectrum data. The modified gravity (MOG) theory resolves the lensing deviation from the standard model and provides an explanation of the CMB and structure growth data.
\end{abstract}

\maketitle

\section{Introduction}

Dark matter was introduced to explain the stable dynamics of galaxies and galaxy clusters and it plays a significant role in the standard $\Lambda$CDM cosmological model. However, dark matter has not been observed in laboratory experiments~\cite{Wimp,Xenon1,Lux}, nor has it been conclusively identified in astrophysical measurements. Therefore, it is important to consider a modified gravitational theory. The observed acceleration of the universe has also complicated the situation by needing a dark energy, either in the form of the cosmological constant vacuum energy or as a modification of GR.

The fully relativistic and covariant modified gravitational (MOG) theory Scalar-Tensor-Vector-Gravity (STVG)~\cite{Moffat,Moffat2,Moffat3,BrownsteinMoffat1,BrownsteinMoffat2,BrownsteinMoffat3,Brownstein,MoffatToth1,MoffatToth2,MoffatToth3,MoffatToth4,MoffatRahvar1,MoffatRahvar2,
MoffatToth5,Roshan1,Roshan2,Roshan3,Roshan4,Roshan5} has been successfully applied to explain the rotation curves of galaxies and the dynamics of galaxy clusters without dark matter. In addition to the attractive gravitational force proportional to the gravitational constant $G$, the theory also has a repulsive gravitational force generated by a massive Proca vector field $\phi_{\mu}$. Before the onset of decoupling at the surface of last scattering about 380,000 years after the big bang, the baryons couple to the photons producing pressure that prevents the overdense instability needed for growth structure. The damping and the dissipative pressure behavior of baryons coupled to photons prevent a GR model without non-baryonic dark matter from fitting the angular acoustical power spectrum. In MOG the CMB Planck data~\cite{Planck3} and the growth of structure to form galaxies and galaxy clusters can be explained assuming that the spin 1 massive dark vector particle, described by the neutral vector field $\phi_\mu$, has a density $\rho_\phi$ that dominates the matter density in the early universe before stars and galaxies are fully formed~\cite{Moffat,Moffat2,Moffat3}. In the late-time present universe the galaxies and galaxy clusters are dominated by baryons with $\rho_b > \rho_\phi$ where $\rho_b$ denotes the density of baryons. The dark vector particle (phion) with gravitational strength coupling to ordinary matter evolves, as the universe expands, into an undetectable particle with negligble mass in the present universe.

Measurements of Cosmic Microwave Background (CMB) anisotropies by the Planck satellite experiment~\cite{Planck1,Planck2,Planck3} have confirmed the expectations of the standard cosmological model based on cold dark matter and the cosmological constant. However, recently Valentino, Melchiorri and Silk~\cite{Silk} have shown that deviations from the standard model have emerged that can give hints of modified gravity. Deviations of the growth of density perturbations predicted by General Relativity (GR) have been reported using a phenomenological parameterization of non-standard metric perturbations~\cite{Planck3}.  Several authors~\cite{Pogosian1,Caldwell1,Simpson,Pogosian2,Dossett} have analyzed the constraints of possible deviations of the evolution of perturbations in the $\Lambda$CDM model by parameterizing the gravitational potentials $\Psi$ and $\Phi$, where $\Psi$ and $\Phi$ denote the Newtonian and curvature potentials, respectively. The constraint analysis performed by Valentino et al., has yielded a result based on the Planck CMB temperature data, parameterized by the parameter $\Sigma_0-1$ that shows a deviation from the standard model. This deviation from the null value expected in GR is about two standard deviations. By including the weak lensing data, the deviation increases to the value: $\Sigma_0-1=0.34^{+0.17}_{-0.14}$. Another anomaly indicated by the Planck data is due to the amplitude of gravitational lensing in the angular spectra. The lensing amplitude parameter $A_{\rm lens}$ is larger than expected in the standard model at the level of $95\%$ c.l.  This anomaly disappears when the modified gravity deviations are taken into account. The parameter $A_{\rm lens}$ is the only one that consistently hints at a tension with standard $\Lambda$CDM. Other possible tensions are hinted at for the amplitude of the r.m.s. density fluctuations parameterized by $\sigma_8$ and the reionization optical depth $\tau$. However, these tensions are much weaker, mainly due to degeneracies between these parameters and $A_{\rm lens}$ and the effective number of relativistic 
neutrinos, $N_{\rm eff}$.

In the following, we will extend the explanation for the growth of structure and the CMB data proposed originally in STVG~\cite{Moffat,Moffat2}. When the density of the neutral, pressureless vector particle associated with the field $\phi_{\mu}$ dominates in the early universe, MOG can lead to a mechanism for the growth of perturbations, solving the cosmological structure problem. The early time domination of the $\rho_\phi$ density determines the acoustical power spectrum in agreement with the Planck 2015 data~\cite{Planck3}. The theory also determines the matter power spectrum in agreement with current large scale galaxy redshift surveys~\cite{MoffatToth3}. 

In MOG the mass of the phion particle is determined by a scalar field $\mu(x)$.  The mass of the phion particle in the early universe is $m_\phi\gg 10^{-28}\,{eV}$, so the phion particle behaves like cold dark matter (CDM). The mass $m_\phi$ dynamically decreases as the universe expands becoming the ultra-light phion particle that is undetectable today. This behavior of the phion particle mass can be explained by a superfluid Bose-Einstein condensate that evolves from the early universe through a phase transition to the present universe in which the Bose-Einstein condensate relaxes and the phion particle mass becomes $m_\phi\sim 10^{-28}$ eV~\cite{Moffat2} . In the late-time present universe, the baryon density $\rho_b$ dominates, $\rho_b > \rho_\phi$, and the MOG theory can fit the galaxy rotation curves and the galaxy cluster dynamics without detectable dark matter. The value of $\mu$ that fits the galaxy rotation curves and the cluster dynamics is $\mu=0.042\,{\rm kpc}^{-1}$~\cite{MoffatRahvar1,MoffatRahvar2}, corresponding to a mass $m_\phi=2.6\times 10^{-28}\,{\rm eV}$.

\section{MOG Field Equations}

The Scalar-Tensor-Vector-Gravity (STVG) field equations are given by~\cite{Moffat,MoffatToth3}:
\begin{equation} 
\label{mog1}
 G_{\mu\nu}+\Lambda g_{\mu\nu}=-8\pi GT_{\mu\nu},
 \end{equation}
 \begin{equation}
\label{mog2}
 \nabla_{\mu}B^{\mu\nu}+ \mu^2\phi^\nu-\frac{\partial V(\phi_\mu)}{\partial\phi_\nu}=4\pi J^\nu,
 \end{equation}
 \begin{equation}
\nabla^\nu\nabla_\nu G=\frac{3}{2}\frac{\nabla^\nu G\nabla_\nu G}{G}-\frac{1}{2}G\frac{\nabla^\nu\mu\nabla_\nu\mu}{\mu^2}-\frac{3}{G}V(G)
+V'(G)-G\frac{V(\mu)}{\mu^2}-\frac{1}{4\pi}(R+2\Lambda),
\label{mog3}
\end{equation}
\begin{equation}
\label{mog4}
\nabla^\nu\nabla_\nu\mu=\frac{\nabla^\nu\mu\nabla_\nu\mu}{\mu}+\frac{\nabla^\nu G\nabla_\nu\mu}{G}-\frac{1}{4\pi}G\mu^3\phi^\mu\phi_\mu-\frac{2}{\mu}V(\mu)+V'(\mu).
\end{equation}
Here, $G_{\mu\nu}=R_{\mu\nu}-\frac{1}{2}g_{\mu\nu}R$, $\nabla_\mu$ denotes the covariant derivative with respect to $g_{\mu\nu}$ and we choose units with $c=1$ and the metric signature $(+1,-1,-1,-1)$. The field $\phi_\mu$ describes a spin 1 neutral vector field, $B_{\mu\nu}=\partial_\mu\phi_\nu-\partial_\nu\phi_\mu$ and $\mu$ is the mass of the vector field. The $V(\phi_\mu), V(G)$ and $V(\mu)$ denote potentials for the $\phi_\mu, G$ and $\mu$ fields and $\Lambda$ is the cosmological constant. 

The total energy-momentum tensor is defined by
\begin{eqnarray} 
T_{\mu\nu}=T^M_{\mu\nu}+T^\phi_{\mu\nu}+T^G_{\mu\nu}+T^\mu_{\mu\nu},
\end{eqnarray} 
where $T^M_{\mu\nu}$ is the energy-momentum tensor for the ordinary matter, and $T^\phi_{\mu\nu}, T^G_{\mu\nu}$ and $T^\mu_{\mu\nu}$ denote the energy-momentum tensors for the $\phi_\mu, G$ and $\mu$ fields. The current density $J^\mu$ is defined by
\begin{equation}
J^\mu=\kappa\rho u^\mu,
\end{equation}
where $\kappa=\sqrt{\alpha G_N}$, $\alpha$ is a parameter, $\rho$ is the total density of matter, $G_N$ is Newton's constant, $u^\mu=dx^\mu/ds$ and $s$ is the proper time along a particle trajectory.

The action for a test particle is
\begin{equation}
S_{\rm tp}=-\biggl( m\int ds+q\int\phi_\mu u^\mu ds\biggr),
\end{equation}
where $m$ and $q=\kappa m=\sqrt{\alpha G_N}m$ are the test particle mass and gravitational charge, respectively, and $\phi_\mu=(\phi_0,\phi_i)\,(i=1,2,3)$. Assuming for simplicity that $V(\phi_\mu)=0$, the weak field spherically symmetric static, point particle solution for $\phi_0(r)$ is obtained from the equation ($\phi_0'=d\phi_0/dr$):
\begin{equation}
\phi_0''+\frac{2}{r}\phi_0'-\mu^2\phi_0=0.
\end{equation}
The solution is given by
\begin{equation}
\label{phisolution}
\phi_0(r)=-Q\frac{\exp(-\mu r)}{r},
\end{equation}
where the gravitational charge $Q=\kappa M=\sqrt{\alpha G_N}M$ and $M$ is the mass of the point particle.

The equation of motion of a test particle is given by
\begin{equation}
\frac{du^\mu}{ds}+{\Gamma^\mu}_{\alpha\beta}u^\alpha u^\beta=\frac{q}{m}{B^\mu}_\nu u_\nu,
\end{equation}
where ${\Gamma^\mu}_{\alpha\beta}$ denote the Christoffel symbols. The equation of motion for a photon is 
\begin{equation}
\frac{du^\mu}{ds}+{\Gamma^\mu}_{\alpha\beta}u^\alpha u^\beta=0.
\end{equation}

We assume that in the slow motion approximation $dr/ds\sim dr/dt$ and $2GM/r\ll 1$, then for the radial acceleration of a test particle we get
\begin{equation}
\frac{d^2r}{dt^2}+\frac{GM}{r^2}=\frac{qQ}{m}\frac{\exp(-\mu r)}{r^2}(1+\mu r).
\end{equation}
For $ qQ/m=\kappa^2M=\alpha G_NM$ and $G=G_N(1+\alpha)$, the modified Newtonian acceleration law for a point particle is given by~\cite{Moffat,MoffatRahvar1}:
\begin{equation}
\label{accelerationlaw}
a(r)=-\frac{G_NM}{r^2}\biggl[1+\alpha-\alpha\exp(-r/r_0)\biggl(1+\frac{r}{r_0}\biggr)\biggr],
\end{equation}
where $r_0=1/\mu$.  The acceleration law can be extended to a distribution of matter:
\begin{equation}
\label{accelerationlaw2}
a({\vec x})=-G_N\int d^3{\vec x}'\frac{\rho({\vec x}')({\vec x}-{\vec x}')}{|{\vec x}-{\vec x}'|^3}
[1+\alpha-\alpha\exp(-\mu|{\vec x}-{\vec x}'|)(1+\mu|{\vec x}-{\vec x}'|)],
\end{equation}
where $\rho(\bf x)$ is the total density of matter. The MOG effective potential for a given density $\rho({\vec x})$ is
\begin{equation}
\Psi_{\rm eff}(\vec{x}) = - G_N \int d^3x'\frac{\rho(\vec{x}')}{|\vec{x}-\vec{x}'|}\Big[1+\alpha -\alpha \exp(-\mu|\vec{x}-\vec{x}'|)\Big].
\label{potential}
\end{equation}

\section{Friedmann Equations and Perturbation Equations}

We base our cosmology on the homogeneous and isotropic Friedmann-Lema\^{i}tre-Robertson-Walker (FLRW) background metric:
\begin{equation}
\label{FLRWmetric}
ds^2=dt^2-a^2(t)\biggl[\frac{dr^2}{1-Kr^2}+r^2(d\theta^2+\sin^2\theta d\phi^2)\biggr],
\end{equation}
where $K=-1,0,+1\,[{\rm length}^{-2}]$ for open, flat and closed universes, respectively. We assume a perfect fluid energy-momentum tensor:
\begin{equation}
T_{\mu\nu}=(\rho+p)u_\mu u_\nu-pg_{\mu\nu},
\end{equation}
where $\rho$ and $p$ are the density and pressure of matter, respectively. We have
\begin{equation}
\rho=\rho_M+\rho_\phi+\rho_G+\rho_\mu+\rho_r,
\end{equation}
where $\rho_M,\rho_\phi,\rho_G$, $\rho_\mu$ and $\rho_r$ denote the density of matter, the neutral vector field $\phi_\mu$ (phion particle), the scalar fields $G$ and $\mu$
and the radiation density, respectively.

Due to the symmetries of the FLRW background spacetime, we have $\phi_0\neq 0,\phi_i=0\,(i=1,2,3)$, $B_{\mu\nu}=0$ and $J^0\neq 0, J^i=0$. We also assume in the following that $V(\phi_\mu)=V(G)=V(\mu)=0$. The MOG Friedmann equations are given by
\begin{equation}
\label{Friedmann1}
\biggl(\frac{\dot a}{a}\biggr)^2+\frac{K}{a^2}=\frac{8\pi G\rho}{3}+\frac{\Lambda}{3}+\frac{\dot a}{a}\frac{\dot G}{G}-\frac{4\pi}{3}\biggl(\frac{{\dot G}^2}{G^2}
+\frac{{\dot\mu}^2}{\mu^2}-\frac{1}{4\pi}G\mu^2\phi_0^2\biggr),
\end{equation}
\begin{equation}
\label{Friedmann2}
\frac{{\ddot a}}{a}=-\frac{4\pi G}{3}(\rho+3p)+\frac{\Lambda}{3}+\frac{1}{2}\frac{\dot a}{a}\frac{\dot G}{G}+\frac{8\pi}{3}\biggl(\frac{{\dot G}^2}{G^2}+\frac{{\dot\mu}^2}{\mu^2}-\frac{1}{4\pi}G\mu^2\phi_0^2\biggr)+\frac{1}{2}\frac{\ddot G}{G}-\frac{{\dot G}^2}{G^2},
\end{equation}
\begin{equation}
\ddot G+3\frac{\dot a}{a}\dot G-\frac{3}{2}\frac{{\dot G}^2}{G}+\frac{1}{2}G\frac{{\dot\mu}^2}{\mu^2}+\frac{1}{8\pi}G\Lambda -\frac{3}{8\pi}G\Biggl(\frac{\ddot a}{a}
+\frac{{\dot a}^2}{a^2}\biggr)=0,
\end{equation}
\begin{equation}
\ddot\mu+3\frac{\dot a}{a}\dot\mu-\frac{{\dot\mu}^2}{\mu}+\frac{1}{4\pi}G\mu^3\phi_0^2=0,
\end{equation}
where $\dot a=\partial a/\partial t$. Moreover, we have the equation:
\begin{equation}
\mu^2\phi_0=4\pi J_0=\kappa\rho.
\end{equation}

In the following, we assume in the early universe cosmology that $\dot G\sim 0$ and ${\dot\mu}\sim 0$ and that the universe is spatially flat, $K=0$.  We obtain the approximate Friedmann equations:
\begin{equation}
\label{Friedmann3}
H^2=\frac{8\pi G\rho}{3}+\frac{\Lambda}{3}+\frac{1}{3}G\mu^2\phi_0^2,
\end{equation}
\begin{equation}
\frac{\ddot a}{a}=-\frac{4\pi G}{3}(\rho+3p)+\frac{\Lambda}{3}-\frac{2}{3}G\mu^2\phi_0^2,
\end{equation}
where $H=\dot a/a$. 

To consider perturbations of the metric and the fields, we write the line element in the conformal Newtonian gauge:
\begin{equation}
ds^2=a^2(\tau)[(1+2\Psi)d\tau^2-(1-2\Phi)dx^idx_i],
\end{equation}
where $\tau$ is the conformal time, $\Psi$ is the Newtonian potential and $\Phi$ is the potential for space curvature.
We have for the Poisson equation for the modified Newtonian potential~\cite{Silk}:
\begin{equation}
k^2\Psi_{\bf k}=-4\pi G_Na^2\zeta\rho\delta,
\end{equation}
where $\zeta$ is a parameter and $\delta=\delta\rho/\rho$ is the relative density perturbation contrast\footnote[1]{We use the parameter $\zeta$ notation instead of $\mu$ used in ref.~\cite{Silk} to avoid confusion with the scalar field $\mu$ notation}.  For the potential $\Phi+\Psi$, the modified Poisson equation is given by
\begin{equation}
k^2(\Psi+\Phi)=-8\pi G_Na^2\Sigma\rho\delta,
\end{equation}
where $\Sigma$ is a parameter and we have $\eta=\Phi/\Psi$ and in GR $\zeta=\Sigma=\eta=1$. 

In MOG the modified Poisson equations are given by
\begin{equation}
\label{PoissonMOG}
k^2\Psi_{\bf k}=-4\pi Ga^2\rho\delta,
\end{equation} 
and
\begin{equation}
k^2(\Psi+\Phi)=-8\pi Ga^2\rho\delta,
\end{equation}
where $G=G_N(1+\alpha)$. In the constraint analysis performed by Valentino et al.,~\cite{Silk}, they find that $\zeta\sim 1$ which corresponds to $G\sim G_N$ in (\ref{PoissonMOG})
and $\alpha\sim 0$.

\section{MOG Perturbations}

We assume that $\rho_\phi$ dominates in the early universe and the $\phi$ field particle describes a neutral dark vector particle with gravitational strength coupling $Q_\phi=\sqrt{\alpha G_N}m_\phi$ to matter, where $m_\phi$ is the mass of the phion particle. At horizon entry until some time after decoupling, $\rho_\phi > \rho_b$, $\rho_\phi >\rho_G$, $\rho_\phi > \rho_\mu$. The first Friedmann equation now becomes
\begin{equation}
\label{Friedmann4}
H^2=\frac{8\pi G\rho_\phi}{3}+\frac{\Lambda}{3}+\frac{1}{3}G\mu^2\phi_0^2.
\end{equation}

In the late-time present universe, baryon matter will dominate and $\rho_\phi <\rho_b$ and MOG gravity will determine the dynamics of galaxies and clusters of galaxies without (detectable) dark matter. The best fit to galaxy rotation curve data and cluster dynamics is obtained with $\alpha=8.89\pm 0.34$ and $\mu=0.042\pm 0.004$~\cite{MoffatRahvar1,MoffatRahvar2}. This value of $\mu$ corresponds to a phion vector particle mass $m_\phi=2.6\times 10^{-28}$ eV. The smallness of the mass in the late-time present universe and the weakness of the gravitational coupling will make the particle undetectable. On the other hand, the dynamics of the scalar field $\mu$ in the STVG field equations will allow for a larger mass $m_\phi$ in the early universe, so that the phion particle can act as dark matter. Thus, $\mu(t)$ will evolve dynamically to a negligible size in the present universe. 

The particle field densities are expressed as the ratios $\Omega_x=8\pi G\rho_x/3H^2$. In particular, we have for the baryon, phion particle and the cosmological constant $\Lambda$:
\begin{equation}
\label{Omegaeqs}
\Omega_b=\frac{8\pi G\rho_b}{3H^2},\quad\Omega_\phi=\frac{8\pi G\rho_\phi}{3H^2},\quad \Omega_{\Lambda}=\frac{\Lambda}{3H^2}.
\end{equation}
At the time of big-bang nucleosynthesis (BBN), we have $\alpha\sim 0$ and $G\sim G_N$, guaranteeing that the production of elements at the time of BBN agrees with observations.  After stars and galaxies are fully formed, $\rho_\phi < \rho_b$ and the MOG non-relativistic acceleration law sets in to explain the rotation curves of galaxies and the dynamics of clusters~\cite{Moffat,MoffatRahvar1,MoffatRahvar2}.

A scenario that can describe the evolution of the phion particle mass $m_\phi$ as the universe expands is to postulate that the phion vector bosons form a cold superfluid of Bose-Einstein condensates before recombination with zero pressure and zero shear viscosity~\cite{Moffat2}. A prediction of statistical mechanics is that a phase transition occurs in an ideal gas of identical bosons when the de Broglie wave length exceeds the mean spacing between bosons. The phion vector bosons in the lowest energy state are stimulated by the presence of other phion bosons to occupy the same state, resulting in a macroscopic quantum-mechanical system. The phion bosons before recombination can collapse into the lowest energy quantum state when the temperature of the phion superfluid is below a critical temperature, $T < T_c$, and form a degenerate Bose-Einstein fluid that does not couple to photons and is not dissipated by photon pressure. The Bose-Einstein condensate superfluid energy-density dominates the energy density before recombination and the density parameter $\Omega_\phi$, the fractional baryon density $\Omega_b$ and the cosmological constant energy-density $\Omega_\Lambda$ can be chosen to fit the Planck angular power spectrum data for the acoustical oscillations in the CMB spectrum~\cite{Planck3}. The quantum phion particle superfluid condensate can clump and form the primordial structure growth for the formation of galaxies. In the late-time universe, the non-relativistic phion matter is dominated by baryon matter and neutral hydrogen and helium. The phion condensate is only subjected to a weak 
gravitational strength force with baryon matter, so that as the universe expands there is little or no decoherence of the phion condensate gas. During the evolution of galaxies and stars 
ordinary baryonic matter begins to dominate inside cores of galaxies and traces light. Cold dark matter halos do not exist in galaxies and clusters of galaxies. A spontaneous symmetry 
breaking leads to a phase transition and the phion Bose-Einstein condensate is relaxed inside galaxies and clusters of galaxies. The effective mass $m_\phi$ of the phion boson is 
significantly reduced, $m_\phi\sim 10^{-28}$ eV, corresponding to a Compton wave length $\lambda_c=1/\mu_\phi=h/m_\phi c\sim 3\times 10^5$ pc.                                                                                                                                                                                                                                                                                                                                                                                                                                                                                                                                                                                                                                                                                                                                 

As the universe expands beyond the time of decoupling, the gravitational attraction between the baryons becomes enhanced as the parameter $\alpha$ increases in size and $G=G_\infty=G_N(1+\alpha)$.  The density $\rho_\phi$ and the increasing enhancement of the size of $G$ deepen the baryon gravitational potential well. After horizon entry and during the era approaching decoupling, the phion mass is $m_\phi\gg 10^{-28}\,{\rm eV}$ corresponding to cold dark matter. The phion mass could be as large as $10^{-6} < m_\phi < 1\,{\rm eV}$. When the earliest stars and galaxies form at about 400 million years after the big bang, $\mu$ undergoes a significant decrease and $m_\phi$ evolves to $m_\phi\sim 10^{-28}\,{\rm eV}$. The Compton wavelength, in natural units, $\lambda_c\sim 1/m_\phi$, of the ultra-light phion particle is of the size of galaxies. 

Because for galaxies and clusters $\rho_\phi <\rho_b$ there is no significant halo of phion particles. Once the combined $\rho_b$ and $\rho_\phi$ density perturbations have grown sufficiently to produce stars and galaxies, then the MOG non-relativistic dynamics for baryons takes over to determine the final evolution and dynamics of galaxies. The matter power spectrum is determined by the transfer function $T_b$ for baryons, which has unit oscillations without dark matter. The oscillations will show up in the calculation of the matter power spectrum. However, the finite size of galaxy survey samples and the associated window function used to produce presently available power spectra mask any such oscillations. Applying a window function to the MOG prediction for the matter power spectrum smooths out the power spectrum curve. The enhanced size of $G=G_N(1+\alpha)$ with $\alpha$ non-zero predicts the right shape for the power spectrum curve, resulting in a fit to the data~\cite{MoffatToth3}. The GR prediction without dark matter and with $G=G_N$ cannot produce the correct magnitude or shape for the matter power spectrum.  In future galaxy surveys which utilize a large enough number of galaxies, with galaxies detected at sufficiently large redshift $z$, and with the use of a sufficiently narrow enough window function, it should be possible to detect any significant oscillations in the matter power spectrum.

\section{Structure Growth and Lensing}

The equation for the density perturbation for the single-component fluid is given by the Jeans equation:
\begin{equation}
\label{deltaperturbation}
\ddot\delta+2H\dot\delta+\biggl(\frac{c_s^2k^2}{a^2}-4\pi G\rho\biggr)\delta=0,
\end{equation}
where we have used the coordinate time $t$ from the FLRW metric (\ref{FLRWmetric}) and $c_s$ is the speed of sound $c_s=\sqrt{dp/d\rho}$. The first term in the parenthesis of (\ref{deltaperturbation}) is due to the adiabatic perturbation pressure contribution $\delta p=c_s^2\delta\rho$. The phion vector particle which couples to ordinary matter with gravitational coupling strength is treated as an almost pressureless particle, so the pressure term in the Jeans equation is absent and $c_s\sim 0$. 
The Jeans length $\lambda_J=2\pi/k_J$ where $k_J=a\sqrt{4\pi G\rho_\phi}/c_s$ is approximately zero and we have for the MOG Jeans equation:
\begin{equation}
\ddot\delta+2H\dot\delta=4\pi G_N(1+\alpha)\rho_\phi\delta_\phi.
\end{equation}

The ratio of the comoving Jeans length to the comoving Hubble length ${\cal H}^{-1}=1/a\sqrt{8\pi G\rho/3}$ is given by ${\cal H}\lambda_J=2\pi\sqrt{2c_s/3}$. Before decoupling the baryon Jeans length is comparable to the Hubble length, so adiabatic baryon perturbations will oscillate before decoupling. The baryon density perturbations begin to grow only after the decoupling time $t=t_{\rm dec}$, because before decoupling the baryon-photon pressure prevents any growth. Without the phion particle density $\rho_\phi$ the baryon density perturbations are too small to produce a growth factor of 1100 needed to generate the observed large-scale structure observed today. However, the phion dark matter particle density perturbations grow after horizon entry and by $t=t_{\rm dec}$ they are significantly greater than the baryon density perturbations. 

Weak lensing of the anisotropies enter as a convolution of the unlensed temperature power spectrum $C_l$ with the lensing potential power spectrum $C_l^{\rm lens}$. The weak lensing parameter $A_{\rm lens}$ is defined as a scaling parameter affecting the lensing potential power spectrum, $C_l^{\rm lens}\rightarrow A_{\rm lens}C_l$, and the standard 
$\Lambda$CDM model has $A_{\rm lens}=1$. The lensing temperature $T^{\rm lens}(\hat n)$ and the unlensed temperature $T(\hat n)$ for weak lensing in cosmology are related by~\cite{Smith}:
\begin{equation}
T^{\rm lens}(\hat n)=T(\hat n +d(\hat n)),
\end{equation}
where $d(\hat n)$ is a vector field representing the deflection angles. The vector $d(\hat n)$ is a pure gradient:
\begin{equation}
d(\hat n)=\nabla\Psi.
\end{equation}
The scalar potential $\Psi$ is given by the line of sight integral for MOG:
\begin{equation}
\Psi(\hat n)=\int_0^{\chi_*}d\chi g'(\chi)[2(1+\alpha)\Psi(\hat n,\chi)],
\end{equation}
where $\chi$ is the comoving distance from the observer, $\chi_*$ is the conformal distance to recombination, $n(\chi)$ denotes the distribution of background sources,
and
\begin{equation}
g'(\chi)=\chi\int_\chi^{\chi(\infty)} d\chi'(1-\chi/\chi')n(\chi').
\end{equation}
This means that the lensing of background galaxies, CMB, or any field of photons, can be influenced by the MOG modification parameter $\alpha$. We can parameterize the lensing parameter $A_{\rm lens}$ by
\begin{equation}
A^{\rm mog}_{\rm lens}=(1+\alpha)A^{\rm st}_{\rm lens},
\end{equation}
where $A^{\rm mog}_{\rm lens}$ and $A^{\rm st}_{\rm lens}$ denote the weak lensing parameter for MOG and for the standard $\Lambda$CDM model, respectively.

Weak lensing measurements of the large-scale structure can capture MOG modifications imposed by $\alpha\neq 0$. We can write the angular power spectrum of lensing convergence $\kappa_c$ as~\cite{Smith,Melchiorri}:
\begin{equation}
C^{\kappa_c}_l=\frac{2}{\pi}\int k^2dk[I^{\kappa_c}(k)]^2P(k),
\end{equation}
where
\begin{equation}
I^{\kappa_c}(k)=\int d\chi g(\chi)[2(1+\alpha)j_l(k\chi)]
\end{equation}
and $j_l(k\chi)$ are the spherical Bessel functions.

We can also probe the measurements of the integrated Sachs-Wolfe effect in MOG~\cite{Caldwell1,Caldwell2}:
\begin{equation}
\frac{\delta T}{T}(\hat n)_{\rm ISW}=\int d\chi[2(1+\alpha)\Psi]_{,\tau}(\hat n,\chi).
\end{equation}
The combined measurements of the CMB with probes of large scale structure, which are sensitive to the modified potential through the Poisson equation, can be used to determine the effects of the MOG parameter $\alpha$.

\section{Data Analysis Results}

Valentino et al.,~\cite{Silk} constrain the cosmological parameters using the full Planck 2015 release on temperature and polarization CMB angular power spectra. The extended $\Lambda$CDM model parameters are the Hubble constant $H_0$, the baryon $\Omega_bh^2$ and cold dark matter $\Omega_ch^2=\Omega_\phi h^2$ energy densities, the primordial amplitude and spectral index of scalar perturbations, $A_s$ and $n_s$, respectively, for the pivot scale $k_0=0.05hMpc^{-1}$, the reionization optical depth $\tau$, the density parameter $\sigma_8$, the neutrino effective number $N_{\rm eff}$, the running of the spectral index $dn_s/d\ln k$ and the lensing parameter $A_{\rm lens}$. The analysis generates constraints on the modified gravity parameters, $\zeta-1, \eta-1$ and $\Sigma_0-1$. The most significant deviations from GR are for the values $\Sigma_0-1=0.36\pm 0.18$ from the Planck TT data and $\Sigma_0-1=0.45^{+0.21}_{-0.17}$ from the Planck $TT +WL\, TT +{\rm lensing}$, where $WL$ denotes the weak lensing data. These results are tabulated in the Tables in ref.~\cite{Silk} for the $68\%$ confidence level on the cosmological parameters, assuming modified gravity and varying the parameters of the extended standard $\Lambda$CDM model.  

The range of values for $\alpha=G/G_N-1$~\cite{Planck2,Planck3,Silk} is $0.2\leq\alpha\leq 0.35$, giving the lensing parameter the range of values:
\begin{equation}
1.2\leq A_{\rm lens}\leq 1.35.
\end{equation}
This result is deduced from the constraint analysis corresponding to a MOG correction to the standard $\Lambda$CDM model for the Poisson equation including weak lensing.  This means that the standard model $A_{\rm lens}$ anomaly disappears when MOG is taken into account. More accurate data on the value of $A_{\rm lens}$ are needed to remove the possibility that the anomaly is due to small systematic experimental errors.

\section{Conclusions}

The STVG modified gravity (MOG) theory is applied to early universe cosmology with an FLRW background geometry. The massive spin 1 vector particle described by the field $\phi_\mu$ is identified as a massive dark vector particle with gravitational strength coupling to matter. Before galaxies are fully formed by structure growth the density $\rho_\phi$ of the $\phi_\mu$ field particles is dominant, $\rho_\phi > \rho_b$.  The particle mass $m_\phi$ in the early universe evolves to a significantly reduced mass, $m_\phi\sim10^{-28}$ eV, in the present late-time universe. The mass $m_\phi$ in the universe today is determined by the best fits of the MOG weak field modified acceleration law to galaxy rotation curves~\cite{MoffatRahvar1} and galaxy cluster dynamics~\cite{MoffatRahvar2} without dark matter. A constraint analysis based on the MOG modified Poisson equation for the gravitational potential $\Psi+\Phi$ with $G=G_N(1+\alpha)$ can explain the perturbative deviations from the $\Lambda$CDM model, discovered by several authors~\cite{Planck2,Planck3,Silk}. Further analysis of the experimental data is needed to verify these MOG deviations from the standard $\Lambda$CDM model. In particular, any small systematic errors in the data need to be uncovered. An important result obtained from the modified gravity is the elimination of the lensing anomaly parameterized by $A_{\rm lens}$ in the standard model. Possible tensions in the determination of the optical depth $\tau$ and the amplitude of the r.m.s. density fluctuations $\sigma_8$ on the scale of $8\,{{\rm Mpc\,h}}^{-1}$ still require further experimental analysis. It is interesting to note that the modified gravity scenario constrains the Hubble constant to be $H_0=68.5\pm 1.1$ at $68\%$ c.l. a value larger than the result $H_0=67.3\pm 0.96$ at $68\%$ c.l. reported by the Planck collaboration using the $\Lambda$CDM model~\cite{Silk}. 

The data obtained from the Planck mission and future data have now reached a degree of accuracy to be able to determine whether the cosmology based on GR is correct, or whether deviations from GR are due to MOG theory. 

\section*{Acknowledgments}

I thank Martin Green, Viktor Toth and Kendrick Smith for helpful discussions. Research at the Perimeter Institute for Theoretical Physics is supported by the Government of Canada through industry Canada and by the Province of Ontario through the Ministry of Research and Innovation (MRI).

\end{document}